\documentclass[prb, superscriptaddress, twocolumn]{revtex4}
\usepackage{simplewick} % To draw Wick's contractions
\usepackage{graphicx}
\usepackage{enumerate}
\usepackage{hyperref}
\usepackage{textcomp}
\usepackage{amsmath}
\usepackage{amssymb}
\usepackage{pgf}
\usepackage{diagbox}
\usepackage{booktabs}
\usepackage{bbold}
\usepackage{color}
\usepackage{bbm}
\usepackage{soul}

\def\ket#1{| #1\rangle}

\newcommand{\quantavg}[1]{\langle #1 \rangle}

\def\ud{\mathrm{d}}
\def\ee{\mathrm{e}}
\newcommand{\nep}{\textrm{e}}

\newcommand{\QA}{\mathrm{\scriptscriptstyle QA}}

\newcommand{\ABC}{\mathrm{\scriptscriptstyle ABC}}

\newcommand{\bomega}{\boldsymbol{\omega}}
\newcommand{\bOmega}{\boldsymbol{\Omega}}

\newcommand{\bbeta}{\boldsymbol{\beta}}
\newcommand{\bgamma}{\boldsymbol{\gamma}}

\newcommand{\x}{{\bf x}}

\newcommand{\opcdag}[1]{{\hat{c}^{\dagger}}_{#1}}
\newcommand{\opc}[1]{{\hat{c}^{\phantom \dagger}}_{#1}}

\newcommand{\Ttime}{\mathrm{T}}

\newcommand{\PauliSigma}{\hat{\sigma}}
\newcommand{\PauliTau}{\hat{\tau}}
\newcommand{\bpaulitau}{{\widehat{\boldsymbol{\tau}}}}

\newcommand{\versorz}{{\hat{\boldsymbol{z}}}}
\newcommand{\versorn}{{\hat{\boldsymbol{n}}}}

\newcommand{\versorb}{{\hat{\boldsymbol{b}}}}

\newcommand{\Texp}{\mathrm{T exp}}
\newcommand{\Tprod}[1]{\prod^{\leftarrow #1}}

\newcommand{\opt}{\mathrm{opt}}

\newcommand{\defects}{\mathrm{def}}

\newcommand{\prob}{\scriptscriptstyle \mathrm{T}}
\newcommand{\driv}{\scriptscriptstyle \mathrm{D}}

\newcommand{\step}{\scriptscriptstyle\mathrm{step}}
\newcommand{\digit}{\scriptscriptstyle\mathrm{digit}}

\newcommand{\Tann}{\tau}
\newcommand{\Ham}{\widehat{H}}
\newcommand{\Hambis}{\widehat{\rm H}}
\newcommand{\Uevol}{\widehat{U}}
\newcommand{\Rrotbb}{\mathbb{R}}
\newcommand{\Rrot}{\mathcal{R}}
\newcommand{\Amat}{\mathbb{A}}
\newcommand{\Pmat}{\mathbb{P}}
\newcommand{\Qmat}{\mathbb{Q}}
\newcommand{\calL}{\mathcal{L}}

\newcommand{\ddef}{\rho_{\defects}}

\newcommand{\ddefdigital}{\rho^{\digit}_{\infty}}
\newcommand{\avgddefstep}[1]{\overline{\rho}_{\defects,{\scriptscriptstyle {#1}}}^{\step}}
\newcommand{\avgddefdigital}[1]{\overline{\rho}_{\defects,{\scriptscriptstyle {#1}}}^{\digit}}

\newcommand{\Ptrot}{\mathrm{P}}
\newcommand{\Uexact}{\Uevol_\QA}

\newcommand{\TermA}{\overline{\Rrotbb[\bomega_{s_{\Ptrot},z}\Tann]}\;}
\newcommand{\TermB}{\Rrotbb[\bomega_{s_{\Ptrot-1},z}\Tann]}
\newcommand{\TermC}{\Rrotbb[\bomega_{s_{\frac{\Ptrot}{2}+1},z}\Tann]}
\newcommand{\TermD}{\; \overline{\Rrotbb[\bomega_{s_{\frac{\Ptrot}{2}},z}\Tann]} \;}
\newcommand{\TermE}{\Rrotbb[\bomega_{s_{\frac{\Ptrot}{2}-1},z}\Tann]}
\newcommand{\TermF}{\Rrotbb[\bomega_{s_1,z} \Tann]}
\newcommand{\TermDots}{\cdots}

\graphicspath{{Figures/}}

\begin{document}

\title{Optimal working point in digitized quantum annealing}

\author{Glen Bigan Mbeng}
\affiliation{SISSA, Via Bonomea 265, I-34136 Trieste, Italy}
\author{Luca Arceci}
\affiliation{SISSA, Via Bonomea 265, I-34136 Trieste, Italy}
%
%\author{Rosario Fazio}
%\affiliation{Abdus Salam ICTP, Strada Costiera 11, 34151 Trieste, Italy}
%\affiliation{NEST, Scuola Normale Superiore \& Istituto Nanoscienze-CNR, I-56126 Pisa, Italy}
%
\author{Giuseppe E. Santoro}
\affiliation{SISSA, Via Bonomea 265, I-34136 Trieste, Italy}
\affiliation{Abdus Salam ICTP, Strada Costiera 11, 34151 Trieste, Italy}
\affiliation{CNR-IOM Democritos National Simulation Center, Via Bonomea 265, I-34136 Trieste, Italy}

\begin{abstract}
We present a study of the digitized Quantum Annealing protocol proposed by R. Barends {\em et al.}, Nature \textbf{534}, 222 (2016). 
Our analysis, performed on the benchmark case of a transverse Ising chain problem, shows that the algorithm has a well defined {\em optimal working point}
for the annealing time $\tau^{\opt}_\Ptrot$ --- scaling as $\tau^{\opt}_\Ptrot\sim \Ptrot$, where $\Ptrot$ is the number of digital Trotter steps --- beyond which 
the residual energy error shoots-up towards the value characteristic of the maximally disordered state. 
We present an analytical analysis for the translationally invariant transverse Ising chain case, but our numerical evidence suggests that this scenario 
is more general, surviving, for instance, the presence of disorder.
\end{abstract}

\maketitle

\section{Introduction} \label{sec:intro}
Quantum annealing (QA)~\cite{Finnila_CPL94,Kadowaki_PRE98,Brooke_SCI99,Santoro_SCI02,Santoro_JPA06},
{\em alias} adiabatic quantum computating~\cite{Farhi_SCI01, Albash_RMP18}, is one of the leading approaches for optimization and
quantum state preparation in currently available quantum devices, broadly belonging to the so-called ``noisy intermediate scale 
quantum technologies'' \cite{Preskill_Quantum2018}. 
The realisation of {\it ad-hoc} quantum hardware implementations based on superconducting flux qubits ~\cite{Harris_PRB10,Johnson_Nat11,Denchev_2016,Lanting_2014,Boixo_2013,Dickson_2013,Boixo_2014}, 
has made QA a field of quite intense research (see Ref.~\onlinecite{Albash_RMP18} for a recent review).
The basic strategy of QA is deeply rooted in the adiabatic theorem of quantum mechanics~\cite{Messiah:book}:
the unknown ground state of some complex target Hamiltonian $\Ham_{\prob}$ is approximated by the state $|\psi(\Tann)\rangle$
obtained through a Schr\"odinger evolution with a suitable $\Ham(t)$ for a sufficiently long annealing time $\tau$, 
starting from a simple initial state $|\psi_0\rangle$. 
In its simplest implementation, one defines $\Ham(t)$ by a linear interpolation
\begin{equation}	\label{H_tot}
	\Ham(t) = s(t) \Ham_{\prob} + (1-s(t)) \Ham_{\driv} \;,
\end{equation}
with $s(0)=0$ and $s(\Tann)=1$, often taken to be a linear function $s(t)=t/\Tann$. 
The driving Hamiltonian $\Ham_{\driv}$ is commonly taken to be a simple transverse field term, 
$\Ham_{\driv}= - \Gamma \sum_{j=1}^N \PauliSigma_j^x$ (for a system of $N$ qubits, $\PauliSigma^x_j$ being a Pauli matrix at site $j$), 
and the initial state $|\psi_0\rangle = |+\rangle^{\otimes N}= 2^{-N/2} (|\! \uparrow\rangle + |\! \downarrow\rangle)^{\otimes N}$ is the ground state of $\Ham_{\driv}$.

As such, the QA approach is intrinsically {\em analog}: it requires constructing the appropriate hardware for a given problem Hamiltonian $\Ham_{\prob}$. 
A recent experimental work by the group of Martinis~\cite{Martinis_Nat16} has advocated a {\em digitized-QA} (dQA), 
in essence a discrete-time Trotter-Suzuki annealing dynamics. 
In the simplest cases, one constructs an approximate state through a depth-$\Ptrot$ digital quantum circuit by splitting the QA Hamiltonian 
into two non-commuting pieces $\Ham_z$ and $\Ham_x$ (involving, say, $\PauliSigma^z$ and $\PauliSigma^x$ Pauli operators):
\begin{equation} \label{eqn:dQA_form}
|\psi_{\Ptrot}\rangle = 
\nep^{-i\beta_\Ptrot \Ham_x} \nep^{-i \gamma_{\Ptrot} \Ham_z} \cdots 
\nep^{-i\beta_1 \Ham_x} \nep^{-i \gamma_{1} \Ham_z} |\psi_0\rangle \;.
\end{equation} 
The dQA parameters $\bbeta = (\beta_1 \cdots \beta_{\Ptrot})$ and $\bgamma=(\gamma_1 \cdots \gamma_{\Ptrot})$ are obtained by applying
an appropriate Suzuki-Trotter splitting to the time-discretized Schr\"odinger evolution operator. 
One great advantage of such a digitized-QA approach is that the different unitaries can be ``constructed''  using universal gates, 
implementing, for instance,  $\PauliSigma^z_i\PauliSigma^z_j$ and $\PauliSigma^x_i\PauliSigma^x_j$ couplings, thus
allowing for the inclusion of {\em non-stoquastic} \cite{Bravyi_QIC08} terms, generally hard to simulate on a classical computer \cite{Troyer_PRL05}. 
A second advantage of a digital implementation is, as argued in Ref.~\onlinecite{Martinis_Nat16}, that the system can in principle be made fully fault-tolerant
 \cite{Kitaev_arXiv1998, Austin_PRA2012}.
 
The dQA form of the state in Eq.~\eqref{eqn:dQA_form} bears an appealing similarity with a hybrid quantum-classical approach known as 
Quantum Approximate Optimization Algorithm (QAOA), introduced by Farhi {\em et al.}~\cite{Farhi_arXiv2014}. 
The non-trivial difference with dQA is that in QAOA the parameters $\bbeta$ and $\bgamma$ are assumed to be {\em variational parameters} of the state
$|\psi_{\Ptrot} (\bbeta,\bgamma)\rangle$, which should be appropriately optimized, via some classical algorithm, by minimizing the expectation value of the 
target Hamiltonian
$E_{\Ptrot}(\bbeta,\bgamma)= \langle \psi_{\Ptrot} (\bbeta,\bgamma) | \Ham_{\prob} |  \psi_{\Ptrot} (\bbeta,\bgamma) \rangle$. 
As such, QAOA is completely insensitive to the possible small gaps encountered by the QA dynamics, and the requirement of adiabaticity is not an issue.
This can bring a variational improvement of the quality of the state $| \psi_{\Ptrot} \rangle$, at the expense of the computational cost for finding the minimum.
%The non-trivial issue with QAOA is how to find good variational parameters, which might increase the computational cost and partially or totally spoil the gain. 

Recently, some of us~\cite{Glen_arXiv2019} have shown that the connection between dQA and QAOA is indeed deeper: 
one can construct, using the QAOA framework, {\em optimal} dQA parameters, which improve over the standard linear schedule $s(t)=t/\Tann$
--- without any need for prior spectral information on the QA Hamiltonian, a notoriously difficult task \cite{Ambainis_arXiv2013, Wolf_Nat2015} ---,
thus realizing within dQA a form of Optimal Quantum Control \cite{Dalessandro2007, Brif_NewJPhys2010, Yang_PRX2017}. 
   
Even without optimization of the parameters, a plain linear-schedule dQA still requires a careful choice of the number of Trotter steps $\Ptrot$,
for a given total annealing time $\Tann$. This is precisely the issue we will analyze in our paper. 
By considering the usual benchmark case of a transverse Ising chain, we show that for any fixed finite $\Ptrot$, a linear-schedule dQA 
has a clear {\em optimal working point} annealing time $\tau^{\opt}_\Ptrot$ --- scaling as $\tau^{\opt}_\Ptrot\sim \Ptrot$ --- beyond which 
the residual energy error shoots-up towards the value characteristic of the maximally disordered state. 
Interestingly, we show that the time-discretization and Trotter errors implied by dQA always lead to a residual energy which is {\em larger} than the
corresponding continuous-time QA, at variance with what is found in Path-Integral Monte Carlo simulated QA \cite{Santoro_SCI02,Heim_SCI15,Mbeng_PRB2019}.   
The scaling behaviour of linear-dQA at the optimal point, however, is identical with the standard Kibble-Zurek scaling \cite{kibble76, zurek85, Polkovnikov_RMP11} 
predicted for continuous-time QA with a linear schedule \cite{Zurek_PRL05,Dziarmaga_PRL05}.  

%We present an analytical proof of this fact for the translationally invariant transverse Ising chain case, but our numerical evidence suggests that this scenario 
%is more general, surviving for instance in the presence of disorder.

The rest of the paper is organized as follows. In Sec.~\ref{sec:model} we introduce the model and the different variants of QA we will study.
In Sec.~\ref{sec:results_dQA} we present our results for digitized-QA with a linear annealing schedule for a translationally invariant quantum Ising chain, discussing 
the presence of the optimal working point $\Tann^{\opt}_{\Ptrot}$ for different values of the Trotter time-steps $\Ptrot$. 
In Sec.~\ref{sec:results_step} we show the effects of time-discretization compared to Trotter digitalization errors. 
In Sec.~\ref{sec:disorder} we discuss the effects of disorder. 
Section~\ref{sec:conclusions} finally, presents our summary and a final discussion of relevant points.  

\section{Model and methods} \label{sec:model}
The class of models considered in Ref.~\onlinecite{Martinis_Nat16} is described by the following QA interpolating Hamiltonian:
\begin{equation}	\label{H_tot}
	\Hambis(s) = s \Ham_{\prob} + (1-s) \Ham_{\driv} \;,
\end{equation}
where $s\in [0,1]$ is a rescaled time, $\Ham_{\driv}= - \Gamma \sum_{j=1}^N \PauliSigma_j^x$ the usual transverse field driving term, 
and $\Ham_{\prob}$ the ``target'' Hamiltonian of which we would like to compute the ground state: 
\begin{equation} \label{eqn:H_P}
	\Ham_{\prob}  = -\sum_{j=1}^N \Big[ J_j^z \PauliSigma_j^z \PauliSigma_{j+1}^z + J_j^x \PauliSigma_j^x \PauliSigma_{j+1}^x  + 
	                                                               B_j^z \PauliSigma_j^z + B_j^x \PauliSigma_j^x \Big] \;.
%	\Ham_{\driv}  &=& - \Gamma \sum_{j=1}^N \PauliSigma_j^x  \;.
\end{equation}
Here $\big( \PauliSigma^x_j, \PauliSigma^y_j, \PauliSigma^z_j \big)$ are Pauli matrices on the $j^\mathrm{th}$ site,
and $N$ the number of sites. 
The transverse field Ising case is recovered for $J_j^x=0$, which can be exactly solved through a Jordan-Wigner mapping~\cite{Lieb_AP61,Young1997}
if we further set also $B_j^z=0$. 
The starting state is assumed, as usual, to be the ground state of the driving Hamiltonian 
%$|\Psi(0)\rangle = \prod_j \frac{1}{\sqrt{2}} (|\!\uparrow\rangle_j+|\!\downarrow\rangle_j)$. 
\begin{equation}
|\psi_0\rangle = 2^{-N/2} (|\!\uparrow\rangle + |\!\downarrow\rangle)^{\otimes N} = |+\rangle^{\otimes N} \;.
\end{equation}
where $|\!\uparrow\rangle$ and $ |\!\downarrow\rangle$ denote the eigenstates of $\PauliSigma^z$, 
while $|+\rangle$ denotes the eigenstate of $\PauliSigma^x$ with eigenvalue $+1$. 
%where $|\mathrm{+x}\rangle_j= (|\!\uparrow\rangle_j+|\!\downarrow\rangle_j)/\sqrt{2}$ is the $+1$ eigenstate of $\PauliSigma^x_j$.

The exact continuous-time QA evolution is given by the time-ordered exponential evolution operator 
\begin{equation}	\label{eqn:Uexact}
	\Uexact(\tau,0) = \Texp\left( -\frac{i}{\hbar} \int_0^\tau \! \ud t \; \Hambis(s(t)) \right) \;.
\end{equation}
%is then digitized by using Trotter break-ups. 
A first step towards full digitalization consists in a time-discretization:
%possible level of ``digitalization'' is to make 
the unitary evolution proceeds in {\em time-steps} where the Hamiltonian is assumed to be constant. 
For the simple case of a linear schedule QA where $s(t)=t/\tau$, with $\tau$ the total annealing time, we divide $\tau$ 
into $\Ptrot$ time-intervals $\Delta t=\tau/\Ptrot$, and write:
\begin{equation}  \label{eq:U_step}
\Uexact(\tau,0) \Longrightarrow
\Uevol_{\step} =  \nep^{-\frac{i\Delta t}{\hbar} \Hambis(s_\Ptrot)} \cdots  \nep^{-\frac{i\Delta t}{\hbar} \Hambis(s_1) } \;, 
%\Tprod{\Ptrot}_{m=1} \nep^{-\frac{i\Delta t}{\hbar} \Ham(s_m) } \;,
\end{equation}
%
%where $\displaystyle\Tprod{\Ptrot}$ denotes a time-ordered product, and 
where $s_m=m/\Ptrot$, with $m=1,\cdots,\Ptrot$. 
To achieve a full digitalization, one must express $\nep^{-\frac{i\Delta t}{\hbar} \Hambis(s_m)}$ in terms of simple quantum gates \cite{Martinis_Nat16}. 
Let us denote by $\Hambis_x(s) = s \Ham_{\prob}^x + (1-s) \Ham_{\driv}$ and $\Hambis_z(s) = s \Ham_{\prob}^z$, 
the terms in $\Hambis(s)$ containing $\PauliSigma^x$ and $\PauliSigma^z$  respectively. 
The simplest possible Trotter splitting of $\Hambis(s)=\Hambis_x(s)+\Hambis_z(s)$ is the $1^{\rm st}$-order one:
\begin{equation} \label{eqn:Trotter_lowest}
\ee^{-i\epsilon \Hambis(s)}  \simeq \ee^{-i\epsilon  \Hambis_x(s)} \ee^{-i\epsilon  \Hambis_z(s)} + O(\epsilon^2) \;,
\end{equation}
which can be easily improved by a symmetrized $2^{\rm nd}$-order splitting:
\begin{equation} \label{eqn:Trotter_symmetrized}
\ee^{-i\epsilon \Hambis(s)}  \simeq  \nep^{-i\frac{\epsilon}{2}  \Hambis_x(s)} \nep^{-i\epsilon  \Hambis_z(s)} 
\nep^{-i\frac{\epsilon}{2}  \Hambis_x(s)} + O(\epsilon^3) \;.
\end{equation}
For ease of notation, and to concentrate on the cases we will actually consider in the numerics, let us set $J_j^x=B_j^x=B_j^z=0$ and define 
\begin{equation}
\left\{ \begin{array}{l} \Ham_z = - \displaystyle \sum_{j=1}^N  J_j^z \PauliSigma_j^z \PauliSigma_{j+1}^z \vspace{2mm} \\
                                  \Ham_x = \Ham_{\driv} = -\Gamma  \displaystyle \sum_{j=1}^N  \PauliSigma_j^x  
\end{array}
\right. \;.
\end{equation}
Then the fully digitized-QA evolution operator can be written as:
\begin{equation} \label{eq:U_digit}
\Uevol_{\digit} = \nep^{-i\beta_\Ptrot \Ham_x} \nep^{-i \gamma_{\Ptrot} \Ham_z} \cdots \nep^{-i\beta_1 \Ham_x} \nep^{-i \gamma_{1} \Ham_z} \;,
\end{equation}
where
\begin{equation} \label{eqn:gamma_m}
\gamma_m =  s_m \frac{\Delta t}{\hbar} \hspace{5mm} \mbox{with} \hspace{5mm} s_m = \frac{m}{\Ptrot} \;,
\end{equation}
while
\begin{subequations}
\begin{align}
	\label{eqn:lowestTrotter}
	\beta_m & = (1-s_m) \frac{\Delta t}{\hbar}   \hspace{17mm} \mbox{\small (1$^{\rm st}$-order Trotter)}  \\
	\label{eqn:symmetricTrotter}
	\beta_m & = \big(1-\frac{s_m+s_{m+1}}{2}\big) \frac{\Delta t}{\hbar}  \hspace{3mm} \mbox{\small (2$^{\rm nd}$-order Trotter)} 
\end{align}
\end{subequations}
with the proviso that we should take $\beta_{\Ptrot}=0$ and neglect an irrelevant phase factor $\nep^{i\Gamma N \Delta t/(2\hbar)}$ 
appearing in the symmetric $2^{\rm nd}$-order splitting. 

We observe that $\Uevol_{\step}$ introduces a {\em time-discretization error}, while  
$\Uevol_{\digit}$ introduces an additional {\em digital (or Trotter) error} \cite{Martinis_Nat16} associated with the non commutativity 
of the quantum operators appearing in $\Uevol_{\step}$.
%Nonetheless the digital protocol has the experimental advantage of requiring only two-qubit gates and was realized in Ref.~\onlinecite{Martinis_Nat16}. 

%\begin{table}
%    \begin{tabular}{|l|l|l|}
%        \hline
%        System                                                                               	& N      	& M \\ \hline
%        \bf{Homogeneous} ($J^z=1, J^x=B^x=B^z=0)$                	& $4$    	& 5  \\ \hline
%        \bf{Homogeneous}  ($J^z=1, J^x= B^x=B^z=0$)               	& $2-6$  	& 3  \\ \hline
%        \bf{Homogeneous} ($J^z=1, J^x= B^x=B^z=0$)               	& $7-9$  	& 2  \\ \hline
%        \bf{Disordered}  								& $3,6$ 	& 5  \\ \hline
%        \bf{Disordered}  								& $7-9$  	& 2  \\ \hline
%    \end{tabular}
%    \caption{Overview of the some of experiments done in Ref~\cite{Martinis_Nat16}. 
%    For the disordered systems the couplings where drawn from independent uniform random distributions such that 
%    $J^z\in[-2,-0.5]\cup[0.5,2]$, $J^x\in[-2,-0.5]\cup[0.5,2]$, $B^x\in[-2,2]$, $B^z\in[-2,2]$.
%    % \glen{I have omitted the experiment with $N=5$ and a single local field on the central spin}.}
%    \label{tab:martinis_istances}
%\end{table}

%The class of models considered above contains exactly solvable cases, for $J_j^x=0$ and $B_j^z=0$, through a Jordan-Wigner mapping
%to a fermionic quadratic Hamiltonian\cite{Lieb_AP61,Young1997}, which we will consider in the following.  
%(In principle, models with $B_j^x\neq 0$ could be Jordan-Wigner solved as well.)
To assess the quality of the annealing, we monitor the density of defects 
created over the ferromagnetic classical Ising ground state~\cite{Dziarmaga_PRL05,Caneva_PRB07}:
\begin{equation}  \label{ddefects}
\ddef(\tau) = \frac{1}{2N} \sum_{j=1}^N \langle \psi(\tau) | (1- \PauliSigma_j^z \PauliSigma_{j+1}^z) | \psi(\tau)\rangle \;,
%\frac{1}{2N} \sum_{j=1}^N \big(1 - \langle \Psi(0) | \Uevol^{\dagger}(\tau) \PauliSigma_j^z \PauliSigma_{j+1}^z \Uevol(\tau) | \Psi(0)\rangle \big)
\end{equation}
where the expectation value is taken over the final state at time $t=\tau$, $|\psi(\tau)\rangle=\Uevol(\tau) |\psi_0\rangle$, calculated 
with different evolution operators. 
More precisely, we will compare: 
\begin{enumerate}
\item The continuous-time QA with a linear-schedule $s(t)=t/\tau$, whose evolution operator is given by Eq~\eqref{eqn:Uexact};
\item The corresponding fully digitized-QA (dQA) with $\Uevol(\tau) =\Uevol_{\digit}$ in Eq.~\eqref{eq:U_digit} with a symmetric Trotter splitting, 
see Eqs.~\eqref{eqn:gamma_m} and \eqref{eqn:symmetricTrotter};
\item The step-discretized QA evolution with $\Uevol(\tau) =\Uevol_{\step}$ in Eq.~\eqref{eq:U_step}, to discriminate digital errors from 
time-discretization errors. 
\end{enumerate}

\section{Results for digitized-QA on Ising chain} \label{sec:results_dQA}
Particularly simple is the case of the translationally invariant transverse Ising model, $J_j^z=J$, with periodic boundary conditions (PBC), 
which reduces to an assembly of $2 \times 2$ independent problems labeled by the momentum $k$, depending on the 
fermionic parity~\cite{Dziarmaga_PRL05}.
A Jordan-Wigner transformation~\cite{JordanWigner_ZPhys1928,Lieb_AP61},
$\PauliSigma^x_j = 1- 2\opcdag{j} \opc{j}$, 
$\PauliSigma^z_j = -(\opc{j} +\opcdag{j})\exp \left(-i\pi\sum_{l=1}^{j-1}\opcdag{l} \opc{l}\right)$, 
%
%\begin{eqnarray} \label{eqn:JordanWigner_map}
%\PauliSigma^x_j &=&   1- 2\opcdag{j} \opc{j} \nonumber \\
%\PauliSigma^z_j & =& -(\opc{j} +\opcdag{j}) \exp \left(-i\pi\sum_{l=1}^{j-1}\opcdag{l} \opc{l}\right)\,,
%\end{eqnarray}
%
maps the spin system to free spinless fermions on a lattice, where $\opcdag{j}$ and $\opc{j}$ respectively 
create and annihilate a fermion at site $j$. 
A Fourier transform can then be used to decompose the system into a set of decoupled two-level systems. 
The transformation is standard, see for instance the appendix in Ref.~\onlinecite{Glen_arXiv2019}. 
The final result can be cast in the following form:
\begin{equation}
\Ham_x = 2\Gamma \sum_{k>0}^{\ABC} \left( \opcdag{k} \opc{k} - \opc{-k} \opcdag{-k} \right) 
\end{equation}
and 
\begin{eqnarray}
\Ham_z &=& -2J  \sum_{k>0}^{\ABC} \bigg( \cos{k} \left( \opcdag{k} \opc{k} - \opc{-k} \opcdag{-k} \right) \nonumber \\
&& \hspace{14mm} + \sin{k}  \left( \opcdag{k} \opcdag{-k} + \opc{-k} \opc{k} \right) \bigg) \;,
\end{eqnarray}
where the sum over $k$ runs over the anti-periodic boundary conditions (ABC) positive wave-vectors given by
$k=\pi(2n - 1)/N$, with $n=1,2,\dots, N/2$, corresponding to the even-fermion-parity sector \cite{Dziarmaga_PRL05}. 
It is convenient to rewrite these Hamiltonians in terms of $N/2$ two-level-systems with effective Pauli matrices
$\bpaulitau_k=(\PauliTau_k^x,\PauliTau_k^y,\PauliTau_k^z)^T$ labelled by the $N/2$ independent $k$-vectors.
With the convention that $|\!\!\uparrow_k\rangle=|0\rangle$ and $|\!\!\downarrow_k\rangle = \opcdag{k} \opcdag{-k} | 0\rangle$,
it is simple to check that the Hamiltonian terms translate as follows:
\begin{equation} \label{eqn:Hx}
\Ham_x = -2\Gamma \sum_{k>0}^{\ABC} \PauliTau^z_k  = -2\Gamma  \sum_{k>0}^{\ABC} \versorz \cdot \bpaulitau_k 
\end{equation}
and 
\begin{equation} \label{eqn:Hz}
\Ham_z = 2J \sum_{k>0}^{\ABC} \big( \cos{k} \, \PauliTau^z_k - \sin{k} \, \PauliTau^x_k \big) =
-2 J \sum_{k>0}^{\ABC} \versorb_k \cdot \bpaulitau_k \;, 
\end{equation}
where we have introduced two unit vectors, $\versorz=(0,0,1)^T$, and $\versorb_k=(\sin k, 0,-\cos k)^T$. 
Observe that, with the previous definitions, the density of defects in Eq.~\eqref{ddefects} simply reads:
\begin{equation} \label{eqn:defects}
\ddef(\Tann) = \frac{1}{2} - \frac{1}{N} \sum_{k>0}^{\ABC}  \versorb_k \cdot \langle \psi_0 | \Uevol^{\dagger}(\tau)  \bpaulitau_k \Uevol(\tau) |\psi_0 \rangle \;.
\end{equation}

\begin{figure}
\includegraphics[width=\columnwidth]{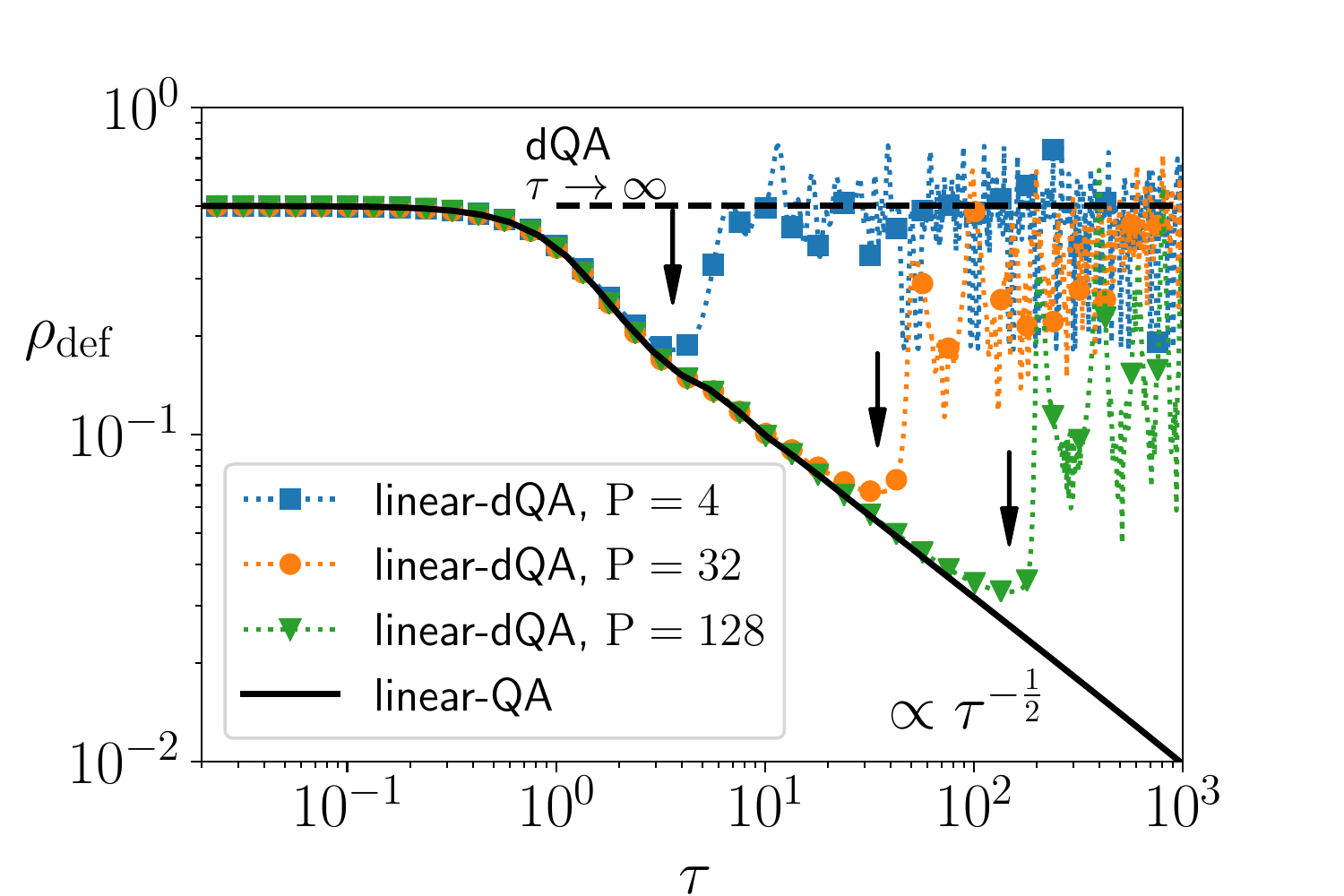}
\caption{Density of defects $\ddef(\tau)$ versus annealing time $\tau$ for an translationally invariant quantum Ising chain with $N=512$ sites
and different fixed values of $\Ptrot$, corresponding to Trotter time-steps of $\Delta t=\tau/\Ptrot$.
The vertical arrows indicate the optimal working points $\Tann^{\opt}_{\Ptrot}$.  
}
\label{fig:ndef_tau_fixP}
\end{figure}

We start by studying how the defect density depends on the total annealing time $\Tann$ for a digitized-QA with a fixed number $\Ptrot$ of quantum gates. 
Henceforth we fix $J=\Gamma$ and measure times in units of $\hbar/J$, equivalently to setting $\hbar=1$ and $J=1$. 
In Fig.~\ref{fig:ndef_tau_fixP} we plot the density of defects $\ddef(\tau)$ versus annealing time $\Tann$ for a translationally invariant Ising chain with $N=512$ sites.
The solid line shows the results of a continuous-time linear-schedule QA --- obtained by numerically solving the time-dependent Schr\"odinger equation ---,
clearly displaying a Kibble-Zurek~\cite{kibble76, zurek85, Polkovnikov_RMP11} (KZ) scaling behaviour \cite{Zurek_PRL05,Dziarmaga_PRL05}. 
The different dotted lines show the results obtained with $\Uevol_{\digit}$, labelled ``linear-dQA'', for three values of $\Ptrot=4, 32, 128$.
%and requiring the application of simple analytical $2\times 2$ unitaries.  
%
As expected, for short and intermediate annealing times ($\Tann < \Ptrot$) time-discritization and digital errors are negligible.  
When $\Tann$ is increased further, the digital protocols show a sharp {\em optimal working point} $\tau^{\opt}_\Ptrot$, which depends on $\Ptrot$
roughly as  $\tau_{\opt}(\Ptrot)\propto \Ptrot$, before $\ddef(\Tann)$ shoots-up towards a rather irregular behaviour  
for $\Tann\to \infty$: we will analytically show below that  $\ddef(\Tann)$ oscillates around the average value $\ddefdigital=1/2$, corresponding to the maximally disordered state. 
The optimal working point is particularly crucial in the dQA protocol, as increasing $\Tann$ beyond $\Tann_{\opt}$ would completely spoil the result and waste resources.  
Incidentally, the large-$\Tann$ irregular behaviour of the digitized-QA data is not a finite-$N$ effect: as explained in Ref.~\onlinecite{Glen_arXiv2019}, the thermodynamic limit
is effectively reached as soon as $N\ge 2\Ptrot+2$. 

To understand the $\Tann\to \infty$ regime at fixed $\Ptrot$, we start by observing that a good way to extract the asymptotic behaviour for
$\ddef(\Tann\to\infty)$ is to calculate an {\em infinite-time average} of $\ddef(\Tann)$:
\begin{equation} \label{eqn:defects_average}
  \overline{\rho}_\defects=\lim_{\Ttime\to \infty}\frac{1}{\Ttime}\int_0^\Ttime \! \ud \Tann \, \ddef(\Tann) \;. 
\end{equation}
For a translationally invariant chain in the thermodynamic limit, where the different $k$-modes become a continuum, exchanging
the time-average integral with the $k$-integral, see Eqs.~\eqref{eqn:defects} and \eqref{eqn:defects_average}, we get:
\begin{equation} \label{eqn:defects_timeaverage}
 \overline{\rho}_\defects =  \frac{1}{2} - \int_0^\pi \! \frac{\ud k}{2\pi} \, \versorb_k \cdot \overline{\quantavg{\bpaulitau_k}_\Tann} 
\end{equation}
where $\overline{\quantavg{\bpaulitau_k}_\Tann}$ is the infinite-time average of:
\begin{equation}
%\overline{\quantavg{\bpaulitau_k}_{\Tann}} = \lim_{\Ttime\to \infty}\frac{1}{\Ttime}\int_0^\Ttime \! \ud \Tann \, 
%\quantavg{\bpaulitau_k}_{\Tann}
\quantavg{\bpaulitau_k}_{\Tann} = \langle \psi_k(0) | \Uevol_k^{\dagger}(\Tann) \bpaulitau_k \Uevol_k(\Tann) |\psi_k(0) \rangle \;.
\end{equation}
The form of the unitary $\Uevol_k(\Tann)$ depends on the QA protocol used. 
The digitized-QA case is particularly simple: 
one can express the action on $\bpaulitau_k$ of the relevant $2\times 2$ unitary operators 
---  $\nep^{i 2\beta_m \versorz \cdot \bpaulitau}$ and   $\nep^{i 2\gamma_m \versorb_k \cdot \bpaulitau}$, in alternation, 
see Eqs.~\eqref{eq:U_digit}, \eqref{eqn:Hx} and \eqref{eqn:Hz} ---
through a product of $3\times 3$ rotation matrices $\Rrot$, by repeatedly using the following Pauli matrix identity:
\begin{equation}
e^{-i\frac{\theta}{2} \versorn \cdot \bpaulitau} \bpaulitau e^{i\frac{\theta}{2} \versorn \cdot \bpaulitau} = \Rrot_{\versorn}(\theta) \bpaulitau \;,
\end{equation}
where $\versorn$ is the axis of rotation, and $\theta$ the rotation angle. 
As a result of that, for the digitized-QA case we end up writing:
\begin{equation}
\overline{\quantavg{\bpaulitau_k}_{\Tann}} = 
\overline{\bigg[ \Tprod{\Ptrot}_{m=1} \Rrot_{\versorz}(4\beta_m)  \Rrot_{\versorb_k}(4\gamma_m) \bigg] } \versorz 
\end{equation}
where $\displaystyle\Tprod{\Ptrot}$ denotes a time-ordered product, and we explicitly used that the initial state is such that 
$\langle \psi_k(0) | \bpaulitau_k |\psi_k(0) \rangle = \versorz$. 
Full details of this analysis can be found in the Appendix.
Concerning the infinite-$\Tann$ average of $\quantavg{\bpaulitau_k}_{\Tann}$, we observe that {\em if} the different 
rotation matrices are {\em uncorrelated}, then we can transform the complicated $\Tann$-average of a product
into a much simpler product of $\Tann$-averages.
It turns out that the fully digital case $\Uevol_{\digit}$ behaves precisely in this way, and one can also show (see Appendix):
\begin{equation}
\int_0^\pi \! \frac{\ud k}{2\pi} \, \versorb_k \cdot \bigg[ \Tprod{\Ptrot}_{m=1} \overline{\Rrot_{\versorz}(4\beta_m)} \; \overline{\Rrot_{\versorb_k}(4\gamma_m)} \bigg] \versorz = 0 \;,
\end{equation} 
which implies that $\overline{\rho}_\defects^{\digit} =  \frac{1}{2}$.
\begin{figure}[t]
\centering
\includegraphics[width=\columnwidth]{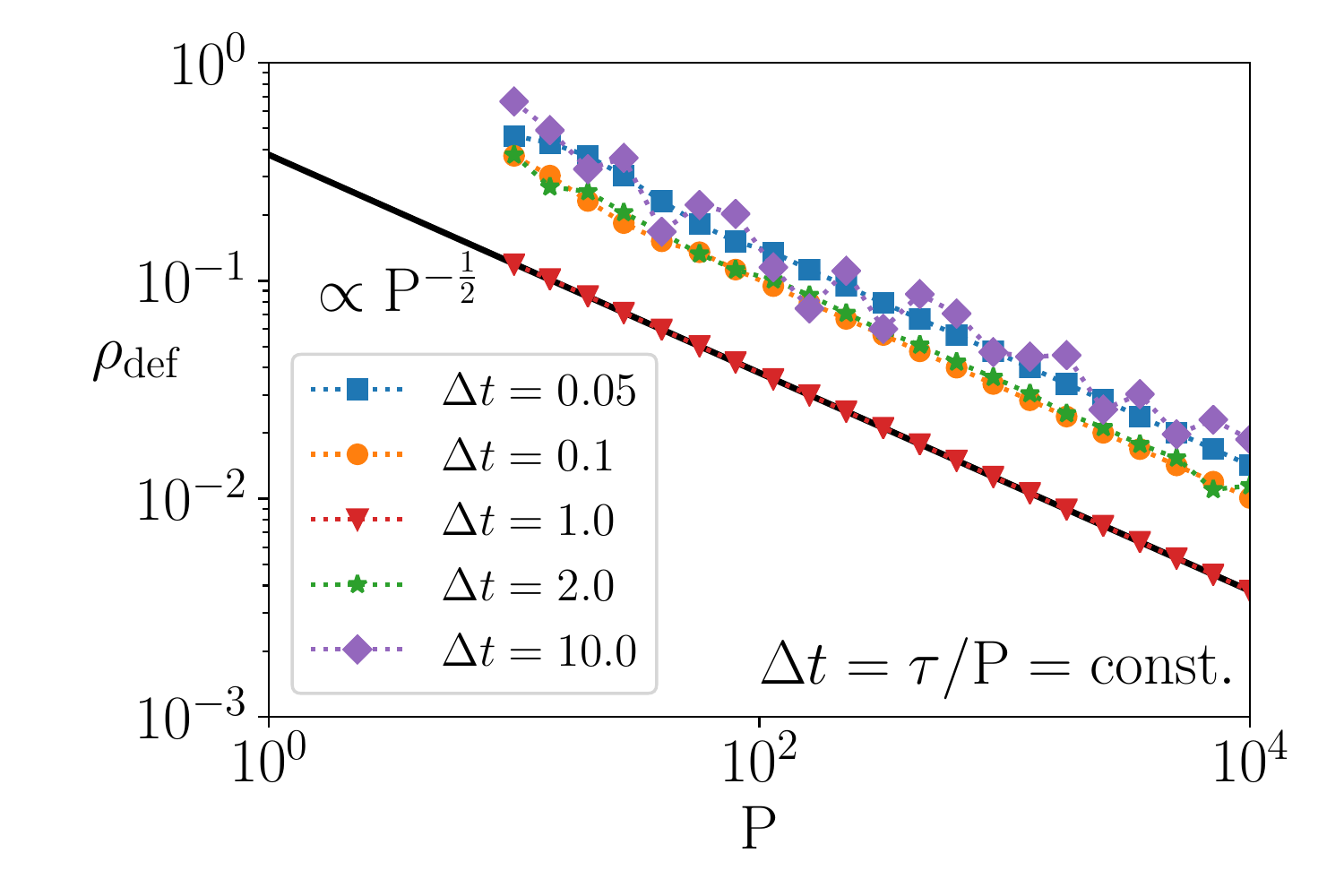}
\includegraphics[width=\columnwidth]{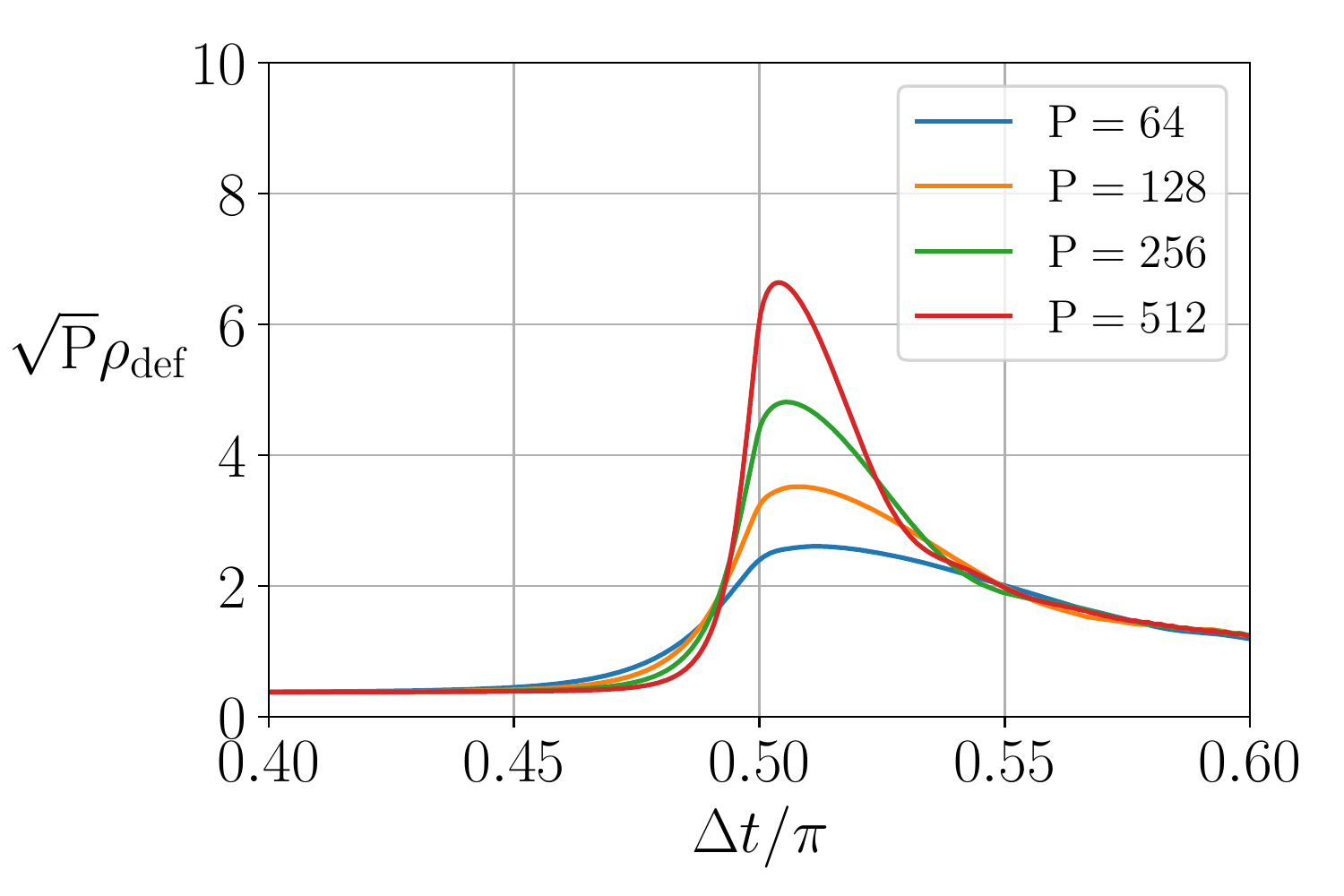}
    \caption{(top) Density of defects $\ddef(\Tann)$ after annealing at several fixed $\Delta t=\Tann/\Ptrot$ as a function of $\Ptrot$ for an ordered Ising 
    chain with $N=1024$. 
    (bottom) Density of defects $\ddef(\Tann)$ after annealing at several fixed $\Ptrot$ as a function of  $\Delta t=\Tann/\Ptrot$ for an translationally invariant Ising 
    chain with $N=1024$. 
}
\label{fig:contant_dt}
\end{figure}

So far we considered digitized-QA at fixed $\Ptrot$. One might ask what happens if one considers results for increasing $\Ptrot$ at constant 
$\Delta t=\Tann/\Ptrot$, corresponding, in the experiment, to applying gates of fixed time-duration. 
Figure~\ref{fig:contant_dt} shows the numerical result we obtained. 
Observe that the best results are obtained when $\Delta t \approx 1$ (in units of $\hbar/J$), in a way that is totally consistent with the optimal working point
shown in Fig.~\eqref{fig:ndef_tau_fixP}, and with the Kibble-Zurek scaling exponent \cite{Zurek_PRL05,Dziarmaga_PRL05}.  
For $\Delta t \ll 1$ the Trotter error is negligibly small but we are wasting resources.  
For $\Delta t \sim 1$ the Trotter error is not small, but the digitized-QA dynamics is very effective and indeed optimal.
For $\Delta t \gtrsim \pi/2$, as shown in Fig.~\ref{fig:contant_dt}(b), there is a sudden increase in the defect density, reflecting the fact that
the digitized-QA dynamics is no longer adiabatic \cite{Glen_arXiv2019, Dranov_JMatPhys1998}: this completely spoils the quality of the results. 

\section{Time-Discretization versus digitalization} \label{sec:results_step}
\begin{figure}
\includegraphics[width=\columnwidth]{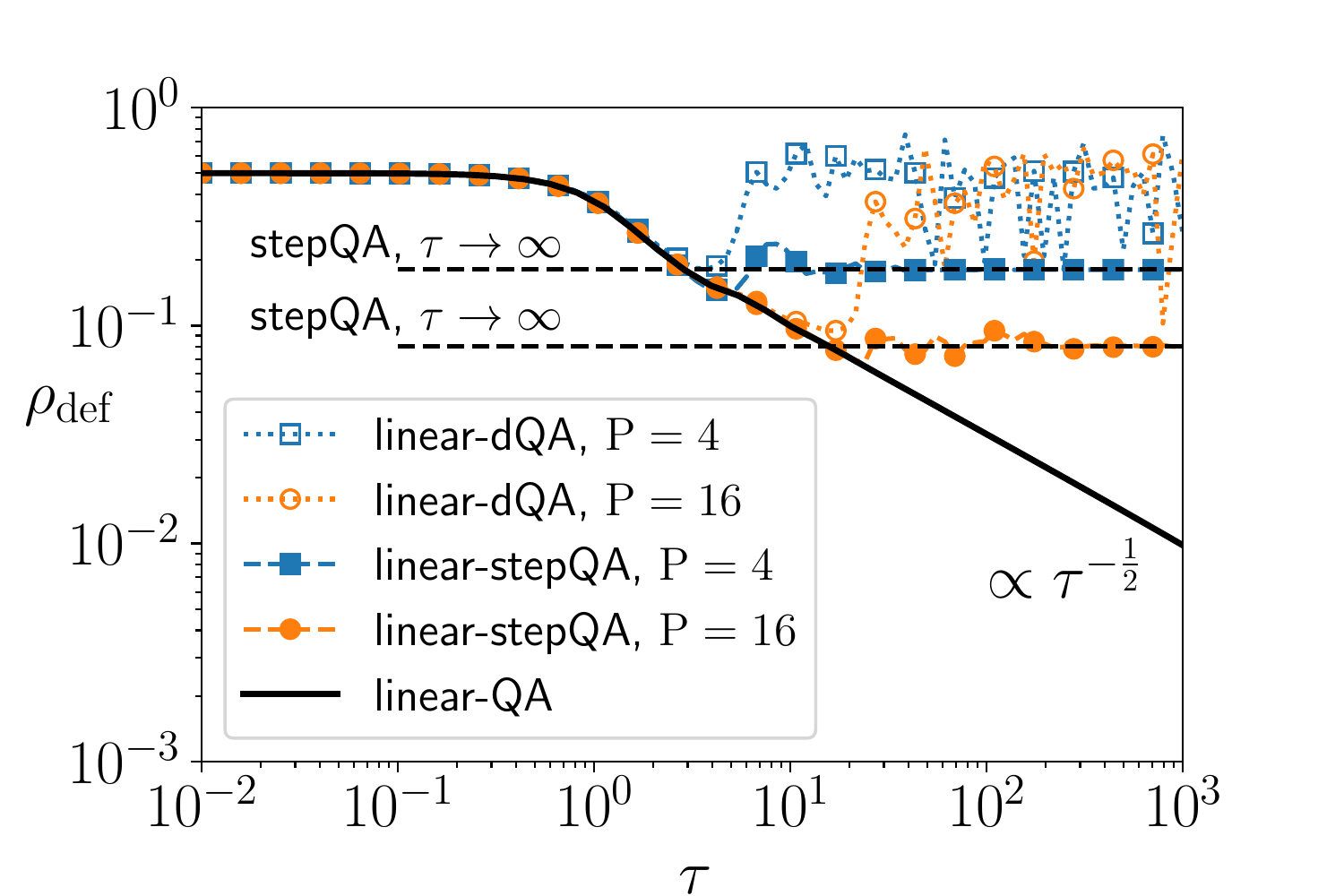}
\caption{Density of defects $\ddef(\Tann)$ versus annealing time $\Tann$ for a translationally invariant quantum Ising chain.
We compare digitized-QA with step-QA results, obtained by a time-discretization of the evolution operator as in Eq.~\eqref{eq:U_step}. 
Here $\Ptrot=4$ and $16$ while $N=1024$. %$16384$, showing negligible size-effects. 
The horizontal solid lines are the analytical predictions for the infinite-time averages in the step-QA case, 
$\overline{\rho}_{\defects,{\scriptscriptstyle \Ptrot=4}}^{\step}=\frac{13}{72}$ and 
$\overline{\rho}_{\defects,{\scriptscriptstyle \Ptrot=16}}^{\step}$.}
\label{fig:ndef_tau_step}
\end{figure}

We now compare the result of a full digitalization with those obtained by a time-discretization of the evolution operator in $\Ptrot$ time steps of $\Delta t=\Tann/\Ptrot$,
which we will denote as step-QA, see Eq.~\eqref{eq:U_step}, with the same linear schedule $s(t)=t/\Tann$. 
Figure~\ref{fig:ndef_tau_step} shows the comparison between digitized-QA and step-QA for the usual translationally invariant Ising chain. 
Notice how the behaviour is very similar for small $\Tann < \Ptrot $ --- indeed almost indistinguishable from the continuous-time QA result.
For larger $\Tann$, however, the behaviour is radically different: $\ddef(\Tann)$ computed with $\Uevol_{\step}$ saturates to a finite plateau 
value which decreases as $\Ptrot$ is increased, at variance with the digitized-QA results: time-discretization and digital errors are clearly 
distinguishable in that regime. It turns out that an analytical determination of the plateau value is still possible even in the step-QA case, 
but the algebra is considerably more involved, because of correlations between the different rotation matrices applied to $\bpaulitau$. 
Without entering into details, reported in the Appendix, we just mention that the calculation of the infinite time average in Eq.~\eqref{eqn:defects_timeaverage}
is reduced, for a given value of $\Ptrot$, to a contour integral over the unit circle $C$ in the complex plane of a rational function
\begin{equation}
\overline{\rho}_{\defects,{\scriptscriptstyle \Ptrot}}^{\step}  =  \frac{1}{2}  +\frac{1}{2\pi i } \oint_C \! \ud z \frac{f_{\scriptscriptstyle \Ptrot}(z)}{g_{\scriptscriptstyle \Ptrot}(z)} 
\end{equation}
where $f_{\scriptscriptstyle \Ptrot}(z)$ and $g_{\scriptscriptstyle \Ptrot}(z)$ are polynomials of the complex variable $z$. 
The integrand (for even $\Ptrot$) has $\Ptrot/2$ $2^{\rm nd}$-order poles given by  the roots of the polynomial $g_{\scriptscriptstyle \Ptrot}(z)$ inside the unit circle $C$, 
located at $z_m = \frac{m}{\Ptrot-m}$ with $m =0,1, \cdots, (\Ptrot/2-1)$.
By calculating explicitly the sum of the residues we find that, for instance,  
$\overline{\rho}_{\defects,{\scriptscriptstyle \Ptrot=2}}^{\step}=\frac{1}{4}$, $\overline{\rho}_{\defects,{\scriptscriptstyle \Ptrot=4}}^{\step}=\frac{13}{72}$, 
$\overline{\rho}_{\defects,{\scriptscriptstyle \Ptrot=6}}^{\step}=\frac{18631}{128000}$, while higher $\Ptrot$ lead to large fractions which we can calculate
with Mathematica. 
As shown in Fig.~\ref{fig:ndef_tau_step}, our analytical prediction matches very well the numerical simulations. 

\section{The effect of disorder} \label{sec:disorder}
\begin{figure}
\includegraphics[width=\columnwidth]{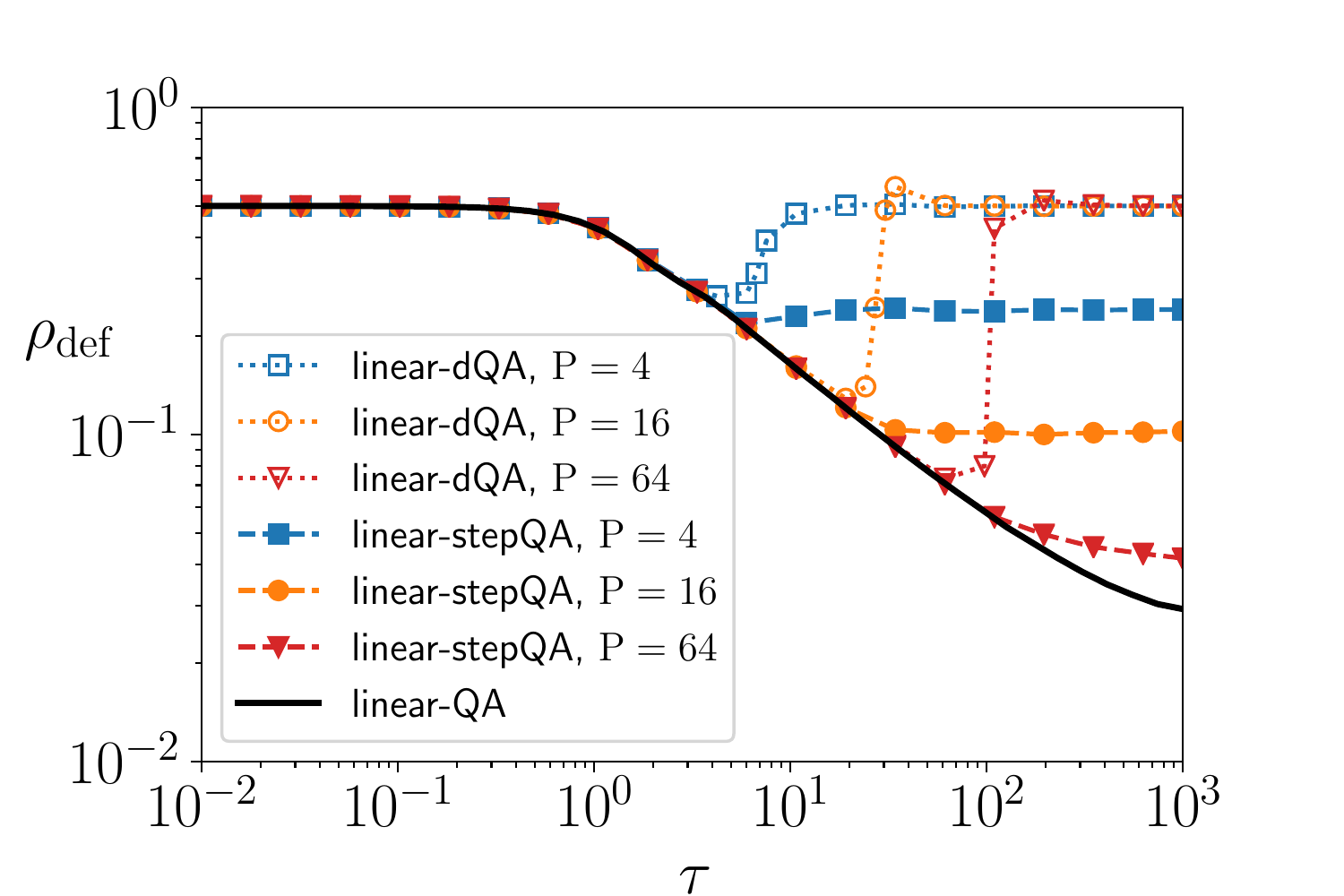}
\caption{Density of defects $\ddef(\Tann)$ versus annealing time $\Tann$ for a disordered quantum Ising chain of length $N=128$, 
averaged over $1000$ realizations of disorder with $J_j^z\in (0,1]$.
}
\label{fig:ndef_tau_fixP_dis}
\end{figure}
The main features we have found so far are not limited to a translationally invariant Ising model. 
In principle, an analysis based on unitary operators can be still pursued in the presence of disorder, as long as one can apply a Jordan-Wigner transformation
obtaining a quadratic fermionic Hamiltonian. The matrices are now $2N\times 2N$ rather than $2\times 2$. 
We have not carried out such analytical calculation. 
However, one can calculate numerically the annealing results for $\ddef(\Tann)$ with the various QA protocols, reported in 
Fig.~\ref{fig:ndef_tau_fixP_dis} for  a disordered chain with $N=128$. 
Here the couplings are taken to be $J^z_j\in (0,1]$, and the results are averaged over $1000$ realizations of the disorder. 
The overall features observed are perfectly consistent with the ordered case results. 
In particular, observe the presence of a sharp optimal working point for the digitized-QA case, where once again $\tau^{\opt}_\Ptrot \propto \Ptrot$.
%
% Pare cosi' dai dati in funzione di Delta t ....

\section{Discussion and conclusions} \label{sec:conclusions}
Summarizing, we have analyzed the effects of Trotter error in digitized-QA for the benchmark case of a transverse field Ising chain, 
highlighting the presence of a sharp optimal working point with $\tau^{\opt}_\Ptrot \propto \Ptrot$, which should be taken care of,
to avoid wasting resources or even spoiling the annealing results with large digital errors. 
The results we have obtained are consistent with the clear minima observed in the experiment, see in particular Fig.~3 of Ref.~\onlinecite{Martinis_Nat16}. 
Interestingly, at the optimal working point the scaling behaviour is precisely consistent with the Kibble-Zurek behavior \cite{Zurek_PRL05,Dziarmaga_PRL05}
seen for continuous-time QA with a linear schedule $s(t)=t/\Tann$. 
We have also analyzed the effect of time-discretization without Trotter splitting, showing that the digital error behaves in a drastically different way for large annealing times.
 
We note, to conclude, that digital errors are seen to lead to a density of defects which is always {\em larger} than that for continuous-time QA.
This is at variance with what is seen when one implements a {\em simulated} QA dynamics by Path-Integral Monte Carlo~\cite{Santoro_SCI02,Heim_SCI15,Mbeng_PRB2019}, 
where digital errors can {\em lower} the density of defects for intermediate annealing times.
It would be extremely interesting to know what digitalization would do in a context of open quantum system dynamics, where the effect of the environment 
on the QA dynamics is duly accounted for. This is an issue worth investigating, which we leave to future studies.

\section*{ACKNOWLEDGMENTS}
We acknowledge fruitful discussions with R. Fazio. 
Research was partly supported by EU Horizon 2020 under ERC-ULTRADISS, Grant Agreement No. 834402.

\bibliography{BiblioQIC,BiblioQAOA,BiblioQA,QIsing} 
%{BiblioQIsing,BiblioQIsing2,BiblioQA,BiblioLNotes,BiblioLZ}
%---------------------------------
%\newpage

\appendix
%---------------------------------
\section{Asymptotic defect density and time-averages} \label{appendix:compute_asymptotics}
We rewrite the system Hamiltonian
\begin{equation}
\Ham(s) = s \Ham_z + (1-s) \Ham_x = - \sum_{k>0}^{\ABC} \hbar \bomega(s, \cos k) \cdot \frac{\bpaulitau_k}{2} \nonumber
\end{equation}
where we defined $\bomega(s,u=\cos k)$ to be the three dimensional vector
\begin{eqnarray}
\hbar \bomega(s,u) &=& 4J s \versorb_k + 4\Gamma (1-s) \versorz \nonumber \\
                               &=& 4(J s \sqrt{1 - u^2}, 0,\Gamma(1 - s) - J s u)^T 
\end{eqnarray}
with $\versorz=(0,0,1)^T$ and $\versorb_k=(\sin k, 0, -\cos k)^T$ as in the main text. 
The unitary gates in Eqs.~\eqref{eq:U_step} and \eqref{eq:U_digit} rotate each pseudo-spin $\bpaulitau$ around a fixed axis, which depends on the momentum $k$
and possibly on $s$. 
In particular, we will use the following Pauli matrix identity:
\begin{equation}
\nep^{-i\frac{\theta}{2}  \versorn \cdot \bpaulitau} \bpaulitau \nep^{+i\frac{\theta}{2} \versorn \cdot \bpaulitau} = \Rrot_{\versorn}(\theta) \bpaulitau \;, 
\end{equation} 
where $\Rrot_{\versorn}(\theta)$ is a $3\times 3$ rotation matrix around the axis $\versorn$ by an angle $\theta$, acting on the pseudo-spin Cartesian components.  
It will be useful, in the following, to have an explicit expression for such a rotation matrix for generic $\versorn$ and $\theta$.
For that purpose, we recast it in the form $\Rrot_{\versorn}(\theta) = \Rrotbb[\bOmega=\theta\versorn]$, where $\Omega=\sqrt{\bOmega \cdot \bOmega} = \theta$
and:
\begin{eqnarray} \label{eqn:rotation}
(\Rrotbb[\bOmega])_{ij}  &=& \frac{\Omega_{i}\Omega_{j}}{\Omega^2} +  \left( \delta_{ij} -  \frac{\Omega_{i}\Omega_{j}}{\Omega^2} \right) \cos \Omega \nonumber \\ 
    				     & & \hspace{10mm} +  \left( \sum_{k=1}^3 \epsilon_{ijk} \frac{\Omega_k}{\Omega} \right) \sin \Omega \;.
%				&=& n_i n_j +  \left( \delta_{ij} -  n_i n_j \right) \cos \theta  + \left( \sum_{k=1}^3 \epsilon_{ijk} n_k \right) \sin \theta \nonumber \;.
\end{eqnarray} 
%
% Ho verificato che questo e' correttamente compatibile con l'espressione in PauliIdentity_Glen.pdf
%

To perform a step-QA dynamics, the typical ingredient needed would be:
\begin{equation}
 \nep^{\frac{i\Delta t}{\hbar} \Ham(s_m)}  \bpaulitau_k  \nep^{-\frac{i \Delta t}{\hbar} \Ham(s_m)} =  \Rrotbb[\bomega(s_m,\cos k)\, \Delta t]  \bpaulitau_k \;.
\end{equation}
In a digitized-QA, the ingredient needed is:
\begin{equation}
 \Uevol^{\dagger}_m \bpaulitau_k  \Uevol_m =  \Rrotbb[4\beta_m\Gamma \versorz ] \, \Rrotbb[4\gamma_mJ \versorb_k ] \bpaulitau_k \;,
\end{equation}
where $\Uevol_m= \nep^{-i\beta_m \Ham_x}  \nep^{-i\gamma_m \Ham_z}$ and one should observe the order of the rotation matrices applied.
Recall now that the density of defects, see Eq.~\eqref{eqn:defects}, is expressed in the thermodynamic limit as:
\begin{equation} \label{eqn:defects_bis}
\ddef(\Tann) = \frac{1}{2} - \int_0^{\pi} \! \frac{\ud k}{2\pi} \, \versorb_k \cdot \langle \psi_0 | \Uevol^{\dagger}(\tau)  \bpaulitau_k \Uevol(\tau) |\psi_0 \rangle \;.
\end{equation}
For the step-QA case the quantum average needed is:
\begin{eqnarray}
    \quantavg{\bpaulitau_k}^{\step}_{\Tann} &=& \quantavg{\psi_0|\Uevol_{\step}^\dagger(\Tann) \bpaulitau_k\Uevol_{\step}(\Tann) |\psi_0} \nonumber \\
    &=& \bigg[ \Tprod{\Ptrot}_{m=1} \Rrotbb[\bomega(s_m,\cos k) \textstyle{\frac{\Tann}{\Ptrot}}] \bigg] \quantavg{\psi_0|\bpaulitau_k|\psi_0} \;, \hspace{3mm}
\end{eqnarray}
where $\displaystyle\Tprod{\Ptrot}$ denotes a time-ordered product, and $\ket{\psi_0}$ is the initial state of the system. 
Since $\ket{\psi_0}$ is chosen to be the ground state of the initial Hamiltonian $\Ham(0) =\Ham_x$ we have:
\begin{equation}
\Ham(0) = -2\Gamma \sum_{k>0}^{\ABC} \versorz \cdot \bpaulitau_k \hspace{1mm} \Longrightarrow \hspace{1mm}
    \quantavg{\psi_0|\bpaulitau_k|\psi_0}= \versorz \;.
\end{equation}
This leads to the equations:
\begin{eqnarray}
   \quantavg{\bpaulitau_k}^{\step}_{\Tann} &=& \bigg[\Tprod{\Ptrot}_{m=1} \Rrotbb[\bomega(s_m,\cos k)  \textstyle{\frac{\Tann}{\Ptrot}} ] \bigg] \versorz \\
   \quantavg{\bpaulitau_k}^{\digit}_{\Tann} &=&  \bigg[ \Tprod{\Ptrot}_{m=1}  
    \Rrotbb[4\beta_m\Gamma \versorz ]  \, \Rrotbb[4\gamma_mJ \versorb_k ] \bigg] \versorz \;.
\end{eqnarray}
It is convenient to extract the asymptotic defect density $\rho_{\defects}(\Tann\to \infty)$ 
from the infinite-time average:
\begin{equation}
  \overline{\rho}_\defects=\lim_{\Ttime\to \infty}\frac{1}{\Ttime}\int_0^\Ttime \! \ud \Tann \, \rho_{\defects}(\Tann) \;.
\end{equation}
In the thermodynamic limit, exchanging the $k$-integral with the time-integral we get:
\begin{equation}  \label{eq:timeavg_ddef}
 \overline{\rho}_\defects =  \frac{1}{2} - \int_0^\pi \! \frac{\ud k}{2\pi} \, \versorb_k \cdot \overline{\quantavg{\bpaulitau_k}_\Tann} \;.
\end{equation}

From now on we will simplify our notation by adopting units such that $\hbar=1$ and $J=1$. We will also take $\Gamma=1$, so that the 
Ising critical point is located at $s_c=\frac{1}{2}$. 
With this choice of units we have that $\bomega(1,\cos k) = 4 \versorb_k$ and $\bomega(0,\cos k)=4 \versorz$. 

\subsection{Step-QA analysis}
For the step-QA evolution we get
\begin{eqnarray}
    \overline{\quantavg{\bpaulitau_k}}^{\step}_{\Tann} &=& \lim_{\Ttime\to \infty}\frac{1}{\Ttime}\int_0^\Ttime \! \ud \Tann \,\quantavg{\bpaulitau_k}^{\step}_{\Tann} \nonumber \\
    &=& \lim_{\Ttime\to \infty}\frac{1}{\Ttime}\int_0^\Ttime \! \ud \Tann \, \bigg[ \Tprod{\Ptrot}_{m=1} \Rrotbb[\bomega(s_m,\cos k) \textstyle{\frac{\Tann}{\Ptrot}} ] \bigg] \versorz \;,   
    \nonumber
\end{eqnarray}
which leads to
\begin{widetext}
\begin{eqnarray}
     \avgddefstep{\Ptrot}&=&  \frac{1}{2} - \frac{1}{8} \int_{-\pi}^\pi \! \frac{\ud k}{2\pi} \, \bomega(1,\cos k) \cdot \overline{\quantavg{\bpaulitau_k}}^{\step}_{\Tann}= \nonumber \\
     &=&  \frac{1}{2} - \frac{1}{32} \oint_C \! \frac{\ud z}{2\pi i z} \, \bomega_{1,z}^T 
     \lim_{\Ttime\to \infty}\frac{1}{\Ttime} \int_0^\Ttime \! \ud \Tann \, \bigg[ \Tprod{\Ptrot}_{m=1} \Rrotbb[\bomega_{s_m,z} \Tann] \bigg]  \bomega_{0,z} \;,   
\end{eqnarray}
\end{widetext}
where we changed variable to $z=e^{ik}$ (so that $\cos k= \frac{z+z^{-1}}{2}$, $\sin k= \frac{z-z^{-1}}{2i}$) and defined:
\begin{equation}
\bomega_{s,z} \stackrel{\rm def}{=} \bomega(s,{\textstyle \frac{z+z^{-1}}{2}}) \;.
\end{equation} 
$C$ denotes the unit circle in the complex plane.
% 
%\RED{\bf Glen, concordi finora con il segno meno davanti e con i fattori, dovuti alla nuova convenzione che introduce certi fattori $4$ in $\bomega$? 
%Il segno meno mi pare fisicamente sensato vista la forma dell'Hamiltoniana finale, il $\langle \bpaulitau \rangle_{\tau}$ deve allinearsi a $\bomega(1,\cos k)$, non trovi?}
%

As we will show in the following, for any positive integer $\Ptrot$, the integrand is a rational function of $z$ and the integral can be evaluated using the residue theorem.
To show this we start by observing that the frequencies involved in the time average are
\begin{equation}
\omega^2(s_m, u) = 16 [1 - 2s_m(1-s_m)(u+1)] \;.
\end{equation}
These are commensurate (indeed identical) only for pairs that are symmetric with respect to $m=\frac{\Ptrot}{2}$ (assuming $\Ptrot$ to be even) 
because $s_m = 1-s_{\Ptrot-m}$, hence $\omega(s_m, u) = \omega(s_{\Ptrot-m}, u)$.
Therefore in computing the time average of the product, we may neglect all correlations except those between the symmetric pairs 
$\Rrotbb[\bomega_{s_m,z}\Tann]$ and $\Rrotbb[\bomega_{s_{\Ptrot-m},z}\Tann]$.
The central matrix $\Rrotbb[\bomega_{s_{\frac{\Ptrot}{2}},z}\Tann]$ and the leftmost matrix $\Rrotbb[\bomega_{s_{\Ptrot},z}\Tann]$ are unpaired and should be averaged separately.
\begin{widetext}
Here is a scheme of the averages we need to perform:
\begin{equation} \label{eqn:scheme}
\contraction[2ex]{\TermA}{\TermB}{\TermDots \TermC \TermD \TermE \TermDots}{\TermF}
\contraction[1ex]{\TermA \TermB \TermDots}{\TermC}{\TermD}{\TermE}
\TermA \TermB \TermDots \TermC \TermD \TermE  \TermDots \TermF \;.
\end{equation}
%
%\begin{equation}
%\overline{\Rrotbb[\bomega_{s_{\Ptrot},z}\Tann]} \; 
%\overbrace{ {\Rrotbb[\bomega_{s_{\Ptrot-1},z}\Tann]} \; \cdots  \overbrace{ \Rrotbb[\bomega_{s_{\frac{\Ptrot}{2}+1},z}\Tann] \; \overline{ \Rrotbb[\bomega_{s_{\frac{\Ptrot}{2}},z}\Tann]} \; \Rrotbb[\bomega_{s_{\frac{\Ptrot}{2}-1},z}\Tann]} \cdots \; {\Rrotbb[\bomega_{s_1,z} \Tann]} } \;.
%\end{equation}
%% {\Rrotbb[\bomega_{s_{x+1,z}\Tann]}         {\Rrotbb[\bomega_{s_{x-1,z}\Tann]}
%%\overbrace{ {P/2+1} \;  \; {P/2-1} }_{} 
\end{widetext}
To exploit this structure of nested averages, it is convenient to recursively contract (performing the time-integration) two rotation matrices sandwiching the central matrix 
at $m=\frac{\Ptrot}{2}$. By doing this, at each step we can independently average over $\Tann$. 
To carry out the contractions, it is useful to define a rescaled vector:
\begin{eqnarray}
    \bOmega_{s,z} &=& z\bomega_{s,z}  \nonumber \\
           &=& 2\left(-i s (z^2-1), 0,  2(1 - s)z - s  (z^2+1)\right)^T  \hspace{7mm}
\end{eqnarray}
and two $3\times 3$ matrices:
\begin{equation}
    \Pmat_{s, z} = \bOmega_{s,z}\bOmega_{s,z}^T \hspace{5mm} \mbox{and} \hspace{5mm} 
    \Qmat_{s, z} = - \bOmega_{s,z}  \wedge \;,
\end{equation}
where $\Pmat_{s,z}$ and $\Qmat_{s,z}$ should be regarded as acting on vectors as follows:
$\Pmat_{s, z} \x =  \bOmega_{s,z} (\bOmega_{s,z} \cdot \x)$ and  $\Qmat_{s, z} \x =  \x \wedge \bOmega_{s,z}$.
%\RED{\bf Glen, non sono sicuro di aver scritto bene l'ultima uguaglianza. Tu mettivi i simbolo $\wedge$ prima del vettore. Intendevi quello che ho scritto sopra?}
%
Notice that $\bOmega_{s,z}$, $\Pmat_{s,z}$ and $\Qmat_{s,z}$ are polynomials in $z$, and they are crucial ingredients 
appearing in the rotation matrix in Eq.~\eqref{eqn:rotation}:
\begin{widetext}
\begin{equation} \label{eqn:rotation_vectorial}
\Rrotbb[\bomega_{s,z}\Tann]  = \frac{1}{\Omega_{s,z}^2} \Pmat_{s,z} +  \left( \mathbb{1} -  \frac{1}{\Omega_{s,z}^2} \Pmat_{s,z} \right) \cos( \omega_{s,z} \Tann )   
    				    +  \frac{1}{\Omega_{s,z}}  \Qmat_{s,z} \sin( \omega_{s,z} \Tann )  \;.
\end{equation}
\end{widetext}
The time-average of the central matrix $\Rrotbb[\bomega_{\frac{1}{2},z} \Tann]$ is therefore given by:
\begin{equation}
\overline{\Rrotbb[\bomega_{\frac{1}{2},z} \Tann]} = \lim_{\Ttime\to \infty} \frac{1}{\Ttime} \int_0^\Ttime \! \ud \Tann \, \Rrotbb[\bomega_{\frac{1}{2},z} \Tann] = \frac{1}{\Omega_{\frac{1}{2},z}^2} \Pmat_{\frac{1}{2},z} 
\end{equation}
since only the first term in Eq.~\eqref{eqn:rotation_vectorial} contributes.
A similar expression holds for the average of the leftmost matrix.
%
%\begin{equation}
%\overline{\Rrotbb[\bomega_{1,z} \Tann]} 
%= \lim_{\Ttime\to \infty} \frac{1}{\Ttime} \int_0^\Ttime \! \ud \Tann \, \Rrotbb[\bomega_{1,z} \Tann] = \frac{1}{\Omega_{1,z}^2} \Pmat_{1,z} \;. 
%\end{equation}
%
To recursively contract the terms as indicated in Eq.~\eqref{eqn:scheme}, we now define a super-operator $\calL_{s,z}$ 
performing the time-integration of two matrices sandwiching a central term 
$\Amat$ (a $3\times 3$ matrix originating from the previous step) as follows:
\begin{widetext}
\begin{equation}
\contraction[1ex]{\calL_{s,z} \Amat \stackrel{\rm def}{=} \Omega_{s, z}^4}{\Rrotbb[\bomega_{1-s,z} \Tann]}{\, \Amat \,}{ \Rrotbb[\bomega_{s,z} \Tann] }
    \calL_{s,z} \Amat \stackrel{\rm def}{=} \Omega_{s, z}^4  \Rrotbb[\bomega_{1-s,z} \Tann] \, \Amat \, \Rrotbb[\bomega_{s,z} \Tann]  
    \stackrel{\rm def}{=} \Omega_{s, z}^4 \lim_{\Ttime\to \infty} \frac{1}{\Ttime} \int_0^\Ttime \! \ud \Tann \,  \Rrotbb[\bomega_{1-s,z} \Tann] \, \Amat \, \Rrotbb[\bomega_{s,z} \Tann] 
\end{equation}
The time-integral can be easily calculated by exploiting the explicit form of the rotation matrices in Eq.~\eqref{eqn:rotation_vectorial}:
\begin{equation}
    \calL_{s,z} \Amat =  \frac{1}{2} \Omega^4_{s, z} \Amat +\frac{1}{2}\Omega^2_{s, z}(\Qmat_{1-s,z} \Amat \Qmat_{s,z} - \Amat \Pmat_{s,z} -\Pmat_{1-s,z} \Amat ) +
    \frac{3}{2} \Pmat_{1-s,z} \Amat \Pmat_{s,z} \;,
\end{equation}
where we used simple trigonometric integrals, such as $\overline{\cos^2(\omega_{s,z} \Tann)}=\overline{\sin^2(\omega_{s,z} \Tann)}=\frac{1}{2}$, 
and the fact that $\Omega_{1-s,z}=\Omega_{s,z}$. 
With this device, we can write an explicit expression for the time-averaged defect density in terms of a time-ordered product of $\frac{\Ptrot}{2}-1$ super-operators
\begin{eqnarray}
    \avgddefstep{\Ptrot} =  \frac{1}{2}  - \frac{1}{32} 
    \oint_C \! \frac{\ud z}{2\pi i z} \, \frac{\bOmega_{1,z}^T\left[\left({\displaystyle \Tprod{\Ptrot-1}_{m=\frac{\Ptrot}{2}+1}} \calL_{1-s_m,z}\right) \Pmat_{\frac{1}{2},z}\right]\bOmega_{0,z}}
                                                              { z^2  \left( {\displaystyle \prod_{m=\frac{\Ptrot}{2}+1}^{\Ptrot-1}} \Omega^4_{s_m, z} \right) \Omega^2_{\frac{1}{2},z} }\;,   
\end{eqnarray}
where the (unpaired) leftmost rotation has been treated separately and did not contribute. 
%
%\commentjoe{ Glen, L'equazione che riporto sotto e' la tua, con le correzioni che mi hai segnalato, ma fattori due diversi a causa delle convenzioni diverse. Ora
%direi che ci siamo, considerato che (coi i miei fattori 2)  $\Omega_{1,z}^2=16z^2$.}.
%\commentglen{Concordo, ci siamo!}
%
%\begin{eqnarray}
%    \avgddefstep{\Ptrot} =  \frac{1}{2}  -
%    \oint_C \! \frac{\ud z}{2\pi i } \, \frac{\bOmega_{1,z}^T \left[\left({\displaystyle \Tprod{\Ptrot-1}_{m=\frac{\Ptrot}{2}+1}} \calL_{s_m,z}\right) \Pmat_{\frac{1}{2},z}\right] \bOmega_{0,z}}
%                                                              { z  \, \Omega_{1,z}^2 \left( {\displaystyle \prod_{m=\frac{\Ptrot}{2}+1}^{\Ptrot-1}} \Omega^4_{s_m, z} \right) \Omega^2_{\frac{1}{2},z} }\;,   
%\end{eqnarray}
%
\end{widetext}

The polynomial appearing in the denominator can be factorized using
\begin{equation}
    \Omega_{s,z}^4 = 2^8 s^2(1-s)^2 z^2\left(z-\frac{1-s}{s}\right)^2 \, \left(z-\frac{s}{1-s}\right)^2 \;.
\end{equation}
Moreover, we have:
\begin{equation}
   \frac{1}{\Omega^2_{\frac{1}{2},z}} \Pmat_{\frac{1}{2},z} = -\frac{1}{4z} \widetilde{\Pmat}_z \;,
\end{equation}
where  $\widetilde{\Pmat}_z = \widetilde{\bOmega}_z \widetilde{\bOmega}_z^T$ and 
$\widetilde{\bOmega}_z= \left( -i (z+1), 0, -(z-1) \right)^T$. 
We therefore rewrite our final expression for  $\avgddefstep{\Ptrot}$ as the complex integral of a rational function of $z$:
%\begin{widetext}
%\begin{eqnarray}
%    \avgddefstep{\Ptrot} =  \frac{1}{2}  + \frac{1}{2^7} 
%    \oint_C \! \frac{\ud z}{2\pi i} \, \frac{\bOmega_{1,z}^T\left[\left({\displaystyle \Tprod{\Ptrot-1}_{m=\frac{\Ptrot}{2}+1}} \calL_{1-s_m,z}\right) \widetilde{\Pmat}_{z} \right]\bOmega_{0,z}}
%                                                              { z^4  \left( {\displaystyle \prod_{m=\frac{\Ptrot}{2}+1}^{\Ptrot-1}} \Omega^4_{s_m, z} \right)} 
%                                            =  \frac{1}{2}  +  \oint_C \! \frac{\ud z}{2\pi i} \, \frac{f_{\scriptscriptstyle \Ptrot}(z)}{g_{\scriptscriptstyle \Ptrot}(z)}  \;,   
%\end{eqnarray}
\begin{equation}
\avgddefstep{\Ptrot} =  \frac{1}{2}  +  \oint_C \! \frac{\ud z}{2\pi i} \, \frac{f_{\scriptscriptstyle \Ptrot}(z)}{g_{\scriptscriptstyle \Ptrot}(z)}  \;, 
\end{equation}
where $f_{\scriptscriptstyle \Ptrot}(z)$ and $ g_{\scriptscriptstyle \Ptrot}(z)$ are polynomials of $z$:
\begin{widetext}
\begin{eqnarray}
   f_{\scriptscriptstyle \Ptrot}(z) &=& \bOmega_{1,z}^T \left[\left(\Tprod{\Ptrot-1}_{m=\frac{\Ptrot}{2}+1} \calL_{1-s_m,z} \right) \widetilde{\Pmat}_{z} \right]\bOmega_{0,z} \\
   g_{\scriptscriptstyle \Ptrot}(z) &=& 2^{4\Ptrot-1} z^{\Ptrot+2}\prod_{m=\frac{\Ptrot}{2}+1}^{\Ptrot-1} \frac{m^2}{\Ptrot^2}\left(1-\frac{m}{\Ptrot}\right)^2 
                         \left(z-\frac{\Ptrot-m}{m}\right)^2 \left(z-\frac{m}{\Ptrot-m}\right)^2 \nonumber \\
                         &=& 2^{4\Ptrot-1} z^{\Ptrot}\prod_{m=\frac{\Ptrot}{2}+1}^{\Ptrot} \frac{m^2}{\Ptrot^2}\left(1-\frac{m}{\Ptrot}\right)^2 
                         \left(z-\frac{\Ptrot-m}{m}\right)^2 \left(z-\frac{m}{\Ptrot-m}\right)^2 \;.
\end{eqnarray}
Here, in the last expression for $g_{\scriptscriptstyle \Ptrot}(z)$ we have re-expressed a factor $z^2$ has an extra term in the product, with $m=\Ptrot$.
This expression for $g_{\scriptscriptstyle \Ptrot}(z)$ shows a number of poles of the rational function which will be used in the calculation of residues.
%\commentjoe{sospetto che il fattore $z^{\Ptrot}$, che tu scrivi anche come un extra $z^2$ dentro al prodotto di $\Ptrot/2$ termini, si cancelli con il numeratore, 
%ma non sono riuscito a dimostrarlo in generale ... Funziona nei 2 casi di cui ho fatto l'algebra ...}
%\commentglen{sembra plausibile ma non credo che sia immediato}

As an example, for the case $\Ptrot=2$ we have $f_{\scriptscriptstyle \Ptrot=2}(z) = 2^4 z^2 (z-1)^2$ and 
$g_{\scriptscriptstyle \Ptrot=2}(z) = 2^7 z^4$, hence:
\begin{eqnarray}
  \overline{\rho}_{\defects,{\scriptscriptstyle \Ptrot=2}}^{\step}  &=&  \frac{1}{2}  +  \oint_C \! \frac{\ud z}{2\pi i} \, \frac{f_2(z)}{g_2(z)} \nonumber \\
    &=&  \frac{1}{2}  +  \oint_C \! \frac{\ud z}{2\pi i} \, \frac{(z-1)^2}{8 z^2} = \frac{1}{4} \;.
\end{eqnarray}
For $\Ptrot=4$ we get:
\begin{eqnarray}
g_{\scriptscriptstyle \Ptrot=4}(z) &=& 9 \cdot 2^7 z^6 (z-{\textstyle \frac{1}{3}})^2 (z-3)^2 \nonumber \\
f_{\scriptscriptstyle \Ptrot=4}(z)  &=& 2^3 z^4 (z-1)^2 ( 9 z^4 - 92 z^3 + 310 z^2 - 92 z + 9) \;, \nonumber
\end{eqnarray}
\end{widetext}
hence:
%Cancelling the common factors (in particular, a factor $z^4$) we can write:
%
\begin{eqnarray}
  \overline{\rho}_{\defects,{\scriptscriptstyle \Ptrot=4}}^{\step}  &=&  \frac{1}{2}  +\frac{1}{2\pi i } \oint_C \! \ud z \frac{f_4(z)}{g_4(z)} = \frac{13}{72} \;. %\nonumber \\
%    &=& \frac{1}{2}  + \frac{1}{2\pi i } \oint_C \! \ud z  \hspace{0.3cm} \frac{(z-1)^2 ( 9 z^4 - 92 z^3 + 310 z^2 - 92 z + 9)}{48 z^2 (z-\frac{1}{3})^2 ( z-3)^2} = \frac{13}{72}  \;.
\end{eqnarray}

More in general, we can use the residue theorem to calculate:
\begin{equation}
    \avgddefstep{\Ptrot}  =  \frac{1}{2}   
    + \sum_{m=0}^ {\frac{\Ptrot}{2}-1} \left.\frac{(z-z_m)[f_{\scriptscriptstyle \Ptrot}(z) - f_{\scriptscriptstyle \Ptrot}(z_m)]}{g_{\scriptscriptstyle \Ptrot}(z)}\right|_{z=z_m}\; ,
\end{equation}
where $z_m = \frac{m}{\Ptrot-m}$ for $m =0,1 \dots (\frac{\Ptrot}{2}-1)$ are the $\frac{\Ptrot}{2}$ (double) roots of the polynomial 
$g_{\scriptscriptstyle \Ptrot}(z)$ lying inside the unitary circle $C$.

For higher values of $\Ptrot$, the polynomials can be calculated with Mathematica, and integrals then evaluated with the residue theorem. 

\subsection{Digitized-QA analysis}
%
%\RED{\bf Glen, rifaccio il conto da zero, anche in considerazione del diverso splitting, ma con le stesse definizioni di prima. 
%Dovresti ricontrollare i conti attentamente, potrei aver scritto cose inesatte.}.\\
%
For the digitized-QA evolution we get:
\begin{eqnarray}
    \overline{\quantavg{\bpaulitau_k}}^{\digit}_{\Tann} &=& \lim_{\Ttime\to \infty}\frac{1}{\Ttime}\int_0^\Ttime \! \ud \Tann \,\quantavg{\bpaulitau_k}^{\digit}_{\Tann} \nonumber \\
    &=& \lim_{\Ttime\to \infty}\frac{1}{\Ttime}\int_0^\Ttime \! \ud \Tann \, \bigg[ \Tprod{\Ptrot}_{m=1}  \Rrotbb[4\beta_m \versorz ]  \, \Rrotbb[4\gamma_m \versorb_k ]  \bigg] 
    \versorz \;, \nonumber   
\end{eqnarray}
which leads, upon the usual change of variables:
\begin{widetext}
\begin{eqnarray}
     \avgddefdigital{\Ptrot}&=&  \frac{1}{2} - \frac{1}{8} \int_{-\pi}^\pi \! \frac{\ud k}{2\pi} \, \bomega(1,\cos k) \cdot \overline{\quantavg{\bpaulitau_k}}^{\digit}_{\Tann}= \nonumber \\
     &=&  \frac{1}{2} - \frac{1}{32} \oint_C \! \frac{\ud z}{2\pi i z} \, \bomega_{1,z}^T 
     \lim_{\Ttime\to \infty}\frac{1}{\Ttime} \int_0^\Ttime \! \ud \Tann \, \bigg[ \Tprod{\Ptrot}_{m=1}  \Rrotbb[ {\textstyle ( 1 - \frac{s_m + s_{m+1}}{2} )} \bomega_{0,z} \Tann ]  \, 
     \Rrotbb[ s_m \bomega_{1,z} \Tann ]  \bigg]  \bomega_{0,z} \;,   
\end{eqnarray}
\end{widetext}
where we made use of the symmetric Trotter splitting and rescaled variables $\Tann/\Ptrot \to \Tann$ in the time-integral.  
Recall we should assume that $s_m=m/\Ptrot$ with $s_{\Ptrot+1}\equiv 1$, in order to make $\beta_{\Ptrot}=0$.  
The frequencies appearing in the various rotation matrices are now all {\em different}, as one can easily verify. 
The time average of the product then becomes a product of time-averages:
\begin{widetext}
\begin{equation}
 \lim_{\Ttime\to \infty}\frac{1}{\Ttime} \int_0^\Ttime \! \ud \Tann \, \bigg[ \Tprod{\Ptrot}_{m=1}  \Rrotbb[ {\textstyle ( 1 - \frac{s_m + s_{m+1}}{2} )} \bomega_{0,z} \Tann ]  \, 
     \Rrotbb[ s_m \bomega_{1,z} \Tann ] \bigg] =  \Big[ \frac{1}{\Omega_{0,z}^2 \Omega_{1,z}^2} \Pmat_{0,z} \Pmat_{1,z} \Big]^{\Ptrot} \;.
\end{equation}
\end{widetext}
where, we recall that:
\begin{eqnarray}
\bOmega_{0,z} &=& z \bomega_{0,z} = 4z \left(0, 0, 1 \right)^T \nonumber \\
\bOmega_{1,z} &=& z \bomega_{1,z} = 2 \left( -i(z^2-1), 0, -(z^2+1) \right)^T \;,
\end{eqnarray}
which imply:
\begin{eqnarray}
\Omega_{0,z}^2 &=& \bOmega_{0,z} \cdot \bOmega_{0,z} = 16 z^2 \nonumber \\
\Omega_{1,z}^2 &=& \bOmega_{1,z} \cdot \bOmega_{1,z} = 16 z^2 \;.
\end{eqnarray}
Since it will be useful in a moment, we also calculate:
\begin{equation}
\bOmega_{1,z} \cdot \bOmega_{0,z} = - 8 z (z^2+1) \;.
\end{equation}
The final expression can therefore be cast in the form:
\begin{eqnarray}
\avgddefdigital{\Ptrot} &=& \frac{1}{2} - \frac{1}{32} \oint_C \! \frac{\ud z}{2\pi i z} \, \frac{\bOmega_{1,z}^T \big( \Pmat_{0,z} \Pmat_{1,z} \big)^{\Ptrot} \bOmega_{0,z} }{z^2 (16z^2)^{2\Ptrot}}
\nonumber \\ 
&=& \frac{1}{2} - \frac{1}{2^{8\Ptrot+5}} \oint_C \! \frac{\ud z}{2\pi i z} \, \frac{ \big( \bOmega_{1,z} \cdot \bOmega_{0,z} \big)^{\Ptrot+1} }{z^{4\Ptrot+2}} \nonumber \\
&=& \frac{1}{2} + \frac{(-1)^\Ptrot}{2^{5\Ptrot+2}}  \oint_C \! \frac{\ud z}{2\pi i} \, \frac{ (z^2+1)^{\Ptrot+1} }{ z^{3\Ptrot+2} } \;.
\end{eqnarray}
We now observe that for even $\Ptrot$ only even powers of $z$ appear inside the integral, and the residue vanishes.
For $\Ptrot > 1$ and odd, the numerator is a polynomial with maximum degree $z^{2\Ptrot+2}$, hence once again the residue vanishes.
The only case in which the integral gives a contribution is for $\Ptrot=1$, where we get:
\begin{equation}
\avgddefdigital{\Ptrot=1} =  \frac{1}{2}  -  \frac{1}{2^{7}} \oint_C \! \frac{\ud z}{2\pi i} \, \frac{ (z^2+1)^2 }{ z^{5} } =  \frac{1}{2}  -  \frac{1}{2^{7}} = \frac{63}{128} \;, \nonumber
\end{equation} 
which is very close to $1/2$ and probably indistinguishable from it, given the fact that there are large fluctuations around the average (which we do not study here).

%---------------------------------
\end{document}